\RequirePackage{lineno}
\documentclass[aps,prb,showpacs, groupedaddress,twocolumn,superscriptaddress]{revtex4}
\usepackage{amsmath,graphicx,latexsym,times,color}
\usepackage{setspace}
\usepackage{hyperref}
\usepackage{array}
\usepackage{titlesec}
\usepackage{physics}

\newcommand{\V}[1]{\ensuremath{\mathbf{#1}}} 

\let\oldtimes\times  
\renewcommand\times{{\oldtimes}}
\newcommand{\Vk}{\V{k}}
\newcommand{\kpt}{{\Vk-point}}
\newcommand{\kpts}{{\Vk-points}}

\begin{document}
	
	\title{Surface states and related quantum interference in \textit{ab initio} electron transport}
	
	\author{Dongzhe Li}
	\email{dongzhe.li@cemes.fr}
	\affiliation{Department of Physics, Technical University of Denmark, DK-2800 Kongens Lyngby, Denmark}
	\affiliation{CEMES, Universit\'e de Toulouse, CNRS, 29 rue Jeanne Marvig, F-31055 Toulouse, France}
	
	\author{Jonas L. Bertelsen}
	\affiliation{Department of Physics, Technical University of Denmark, DK-2800 Kongens Lyngby, Denmark}
	\affiliation{Center for Nanostructured Graphene, Technical University of Denmark, DK-2800 Kongens Lyngby, Denmark}
	
	\author{Nick Papior}
	\affiliation{DTU Computing Center, Department of Applied Mathematics and Computer Science, Technical University of Denmark, DK-2800 Kongens Lyngby, Denmark}
	
	\author{Alexander Smogunov}
	\affiliation{SPEC, CEA, CNRS, Universit\'e Paris-Saclay, CEA Saclay, F-91191 Gif-sur-Yvette, France}
	
	\author{Mads Brandbyge}
	\email{mabr@dtu.dk}
	\affiliation{Department of Physics, Technical University of Denmark, DK-2800 Kongens Lyngby, Denmark}
\affiliation{Center for Nanostructured Graphene, Technical University of Denmark, DK-2800 Kongens Lyngby, Denmark}
	
	\date{\today}
	
	\begin{abstract}
		Shockley surface states (SS) have attracted much attention due to their role in various physical phenomena occurring at surfaces. It is also clear from experiments that they can play an important role in electron transport. However, accurate incorporation of surface states in $\textit{ab initio}$ quantum transport simulations remains still an unresolved problem. Here we go beyond the state-of-the-art non-equilibrium Green's function formalism through the evaluation of the self-energy in real-space, enabling electron transport without using artificial periodic in-plane conditions. We demonstrate the method on three representative examples based on Au(111): a clean surface, a metallic nanocontact, and a single-molecule junction. We show that SS can contribute more than 30\% of the electron transport near the Fermi energy. A significant and robust transmission drop is observed at the SS band edge due to quantum interference in both metallic and molecular junctions, in good agreement with experimental measurements. The origin of this interference phenomenon is attributed to the coupling between bulk and SS transport channels and it is reproduced and understood by tight-binding model. Furthermore, our method predicts much better quantized conductance for metallic nanocontacts. 
	\end{abstract}
	
	\renewcommand{\vec}[1]{\mathbf{#1}}
	
	\maketitle

\section{INTRODUCTION}	

	At the surfaces of noble metals lacking inversion symmetry, the Shockley surface states (SS) appear in the projected band-gap of bulk states \cite{davison1996basic}. These SS are nearly free-electron model systems due to their confinement to the surface, and thus have attracted significant attention for the investigation of fundamental many-body effects in solids, in particular by angle-resolved photoemission spectroscopy \cite{Kevan1987,Eiguren2002,Sirotti2014}, and scanning tunneling microscopy (STM) \cite{berndt1999surface,Jiutao1998,Kliewer2000,Reinert2001}. Crommie \textit{et al.} \cite{crommie1993} formed quantum corrals on a metal surface where the SS create the wave phenomena. Many transport measurements show that SS are actively involved in adsorption, and chemical processes on surfaces \cite{donath1995electronic,ren2019k}. The manipulation of SS can influence the molecule-metal interaction, altering the Kondo temperature \cite{Iancu2006,Granet2020}, while adsorption of single adatoms can induce further localization of the 2D Shockley SS at the metal surfaces \cite{Madhavan2001,Olsson2004,Merino2004,Wahl2005,Limot2005,Vitali2008}. Moreover, tunneling between molecular orbitals and the SS is shown to produce negative differential conductance in metal-organic junctions \cite{Heinrich2011}. These experimental observations invoke several fundamental questions to be addressed: Can we describe SS accurately in transport calculations? What is the role of SS in electron transport and how can one use them to tune electrical conductance? 
	
\begin{figure}[!t]
		\centering
		\includegraphics[width=1.0\linewidth]{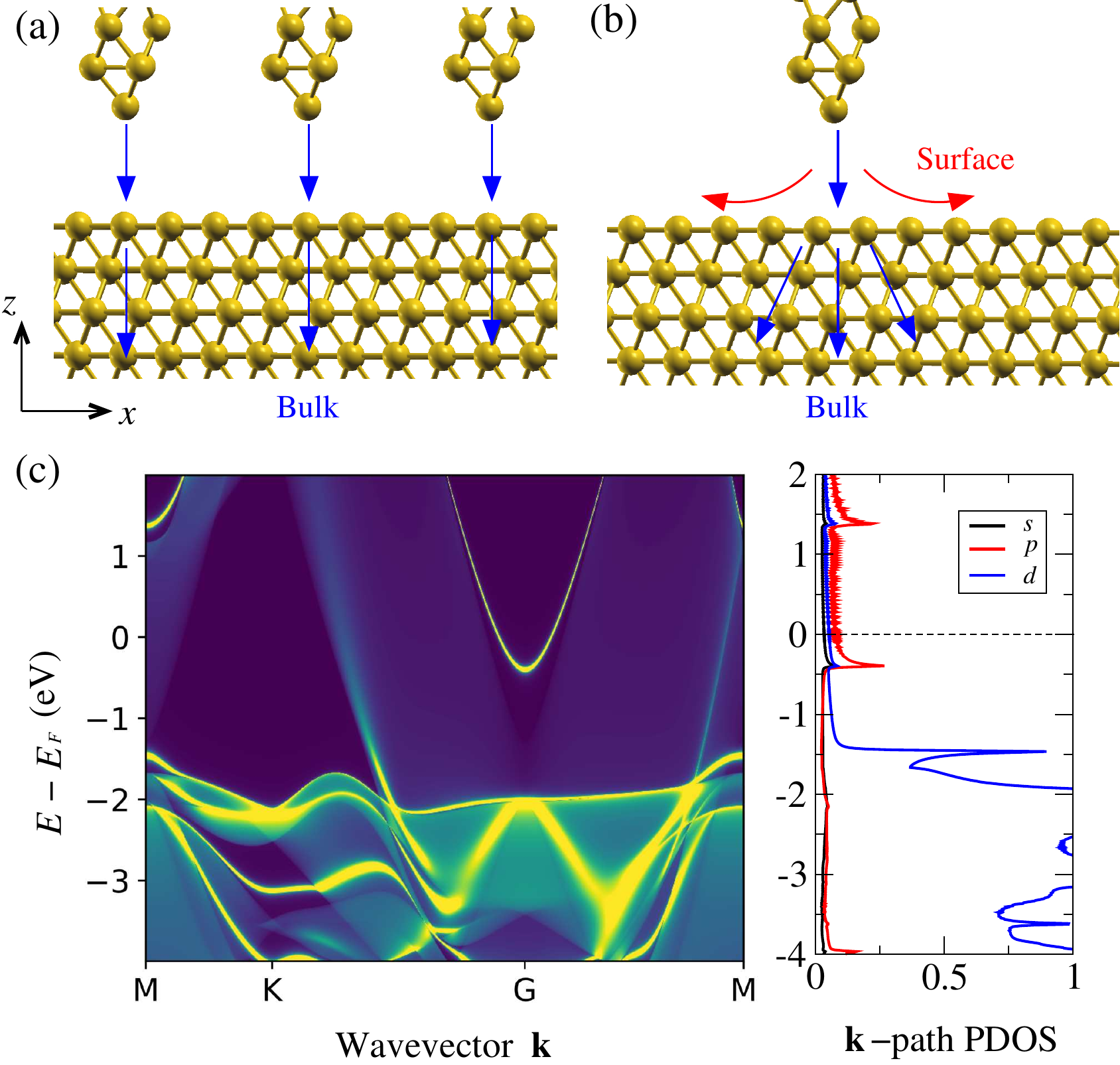}
		\caption{\label{band-structure}
			(a) Supercell approach: Atomic-scale junctions with in-plane periodic boundary conditions (PBCs), (b) RSSE method: Single atomic-scale junctions without PBCs, (c) Clean Au(111) surface: Band structure projected on the surface atom (left) and corresponding PDOS along the $\V{k}$-path MKGM (right).}
	\end{figure}

	During the last two decades, the combination of density functional theory (DFT) and non-equilibrium Green's functions (NEGF) theory has been recognized as the most powerful computational tool with predictive power. Currently, there are mainly two different DFT+NEGF methodologies being applied. The first one is based on supercell approach as implemented in \textsc{TranSiesta} \cite{Brandbyge-2002,papior2017improvements}, \textsc{Nanodcal} \cite{Taylor2001}, \textsc{QuantumATK} \cite{smidstrup2019}, and many other codes \cite{Choi1999,Nardelli1999, Wortmann2002,Smogunov2004,Wang2005,Rocha2006,Pecchia_2008,saha2009first,Ozaki2010,verzijl2012dft,Chen2012,ferrer2014gollum,refaely2019first,takaki2020sake,Meng2021}. They typically remove periodic boundary conditions (PBCs) in the transport direction using the recursive surface self-energy method \cite{sancho1985} while for the directions transverse to the transport, the PBCs are retained, and corresponding \kpts\ are used. This means that the atomic-scale junction under investigation, such as an atomic contact, an adatom, or a molecular junction, as well as the connecting electrodes, such as a tip, are repeated in the transverse direction, as shown in Fig.~\ref{band-structure}a. In this setup, the only current-carrying asymptotic states are the bulk Bloch states propagating deep into the electrodes. Conduction channels related to SS, propagating along the surface (red arrows in Fig.~\ref{band-structure}b), are missing. The second approach is a cluster-based method \cite{jacob2011,bagrets2013,Frank2015} which does not capture the surface state properly due to the finite size of the cluster. Another particular cluster-based approach developed by F. Pauly \textit{et al.} \cite{pauly2008}, a kind of mixed cluster-supercell method, which in principle could include the SS. However, the SS and their contribution to the total self-energy have never been studied with their approach. Moreover, due to its inherent need for large clusters, the SS is approximated, to our knowledge.
	
	Here, we demonstrate the proper incorporation of surface states' contribution in $\textit{ab initio}$ electron transport, going beyond the current state-of-the-art DFT+NEGF framework by removing all PBCs. This is accomplished by calculating the real-space self-energy (RSSE) via the \Vk-averaged Green's function. We show that the SS contributes more than 30\% of the electron transport near the Fermi energy ($E_F$) on noble metal surfaces, resulting in a step-like increase of transmission function. Furthermore, a pronounced transmission drop is observed just below $E_F$ due to quantum interference (QI) when the surface potential is perturbed (by e.g., an STM tip, an adatom, or a molecule). These results are in a good agreement with experimental reports \cite{Madhavan2001,Hollen_2015,Olsson2004, Limot2005}. The step- or drop-like features observed in transmission at the SS band edge and their interpretation is the main goal of this manuscript.
	
\section{METHOD}
	
	The DFT+NEGF transport calculations within the local density approximation (LDA) \cite{Perdew1981} were carried out using \textsc{TranSiesta} \cite{Brandbyge-2002,papior2017improvements}, and the post-processing codes \textsc{TBtrans}, and \textsc{sisl} \cite{zerothi_sisl}. The transmission function is calculated using the NEGF approach as
\begin{equation}
\label{eq0}
T(E)=\Trace[\V{\Gamma_{\text{L}}}\V{G}\V{\Gamma_{\text{R}}}\V{G}^{\dagger}]
\end{equation}
where $\V{G}$ is the retarded Green's function, and $\V{\Gamma_{\text{L}}}$/$\V{\Gamma_{\text{R}}}$ are matrices describing the coupling to the left/right semi-infinite electrodes including surface states. The ballistic conductance is evaluated from the corresponding electron transmission function at $E_F$ using the Landauer formula \cite{blanter2000}, $G=G_0T(E_F)$, where $G_0 = 2e^2/h$ is the conductance quantum. We compare our RSSE method with the supercell approach, where the open boundary condition is applied in the out-of-surface transport direction $z$ while PBCs are used in the surface plane $xy$. The RSSE removes all PBCs by evaluating the ``surrounding/embedding'' self-energy corresponding to the surface supercell in real-space, which can be outlined as
\begin{equation}
\label{eq1}
\V{\Sigma}^{\mathrm{RSSE}}(E)=\V{S}^{\mathrm{RS}}(E+i\eta)-\V{H}^{\mathrm{RS}}-\bigg[ \sum_{\V{k}}\V{G_k}(E) \bigg]^{-1}\,
\end{equation}
Here $\V{G}_\Vk(E)$ is the Green's function for a given \kpt, $\V{H}^{\mathrm{RS}}$ and $\V{S}^{\mathrm{RS}}$ are the real-space representation of Hamiltonian and overlap matrices for an infinitely large system projected on the real-space supercell. The self-energy $\V{\Sigma}^{\text{RSSE}}$ is only nonzero on the boundary of the supercell, where orbitals have matrix elements with the states outside the supercell. Since the self-energy describes a perfect, semi-infinite bulk structure with the single supercell missing, we may reuse this self-energy for any perturbed supercell of the same size containing for example an adatom, vacancy, adsorbate, and STM-tip. All states in the supercell which originate from bulk scattering states are represented in $\V{G_k}(E)$. However, the surface states are orthogonal to these and thus appear as poles in $\V{G_k}(E)$. In order to include them, we used a small imaginary part/broadening $\eta=2\,\mathrm{meV}$, and a correspondingly dense $\V{k}$-points of 400 $\times$ 400 $\V{k}$-mesh for the evaluation of real-space quantities in Eq.~\eqref{eq1}. The implementation details are described in our recent work \cite{Papior2019} where the method was applied only to two-dimensional (2D) systems, here we show its importance in three-dimensional (3D) systems of atomic-scale junctions with a particular focus on SS. We refer to Appendix A for details on the computational parameters and $\V{k}$ points convergence test is presented in Appendix B.

\section{RESULTS}
	
	We start our discussion with the band structure of the clean Au(111) surface. Instead of the slab model \cite{forster2007,Marathe2012,Requist2015}, we use a single surface connected to infinitely many layers. This is done by calculating the $\V{G}(E)$ for six Au(111) layers of which the bottom three layers connect via the self-energy $\V{\Sigma}(E)$ to an infinite substrate, thus removing the PBCs in this direction, but maintaining transverse \Vk-dependence:
	\begin{equation}
	\label{eq2}
	\V{G_k}(E)=\left[(E+i\eta)\V{S_k}-\V{H_k} - \V{\Sigma_k}(E)\right]^{-1}
	\end{equation}
	where $\V{H_k}$ is the Hamiltonian and $\V{S_k}$ the overlap matrix of the surface layers in \Vk-space. We plot in Fig.~\ref{band-structure}c the band structure of Au(111) surface projected on the surface atom, where the well-known band gap and SS at the $\Gamma$ point are evident. The binding energy at $\Gamma$ is about $0.4\,\mathrm{eV}$, in good agreement with previous \emph{ab initio} calculations \cite{Marathe2012,Sirotti2014} and experimental measurements \cite{Andreev2004,forster2007,ifmmode2012}. From the $\V{k}$-path PDOS, we further identify that the SS are mainly from $p$ orbitals.
	
	Having reproduced the electronic structure of SS of a clean Au(111) we pass now to their effect on the transport properties of a point contact as shown in Fig.~\ref{band-structure}b. Standard DFT+NEGF simulations based on the supercell approach are problematic in two ways: i) the simulations have a periodic array of junctions including the repetition of the STM probe tip; such an array of junctions will have unphysical interactions and requires extra care with the convergence of $\V{k}$ points and supercell size \cite{Thygesen2005} and ii) more importantly, the simulations are unable to capture the surface state contribution to the transport. Our RSSE method solves both problems. In the following, we compare RSSE with the supercell approach for three representative examples based on Au(111): a clean surface, a metallic nanocontact, and a single-molecule junction.

\begin{figure}[!t]
	\centering
	\includegraphics[width=1.0\linewidth]{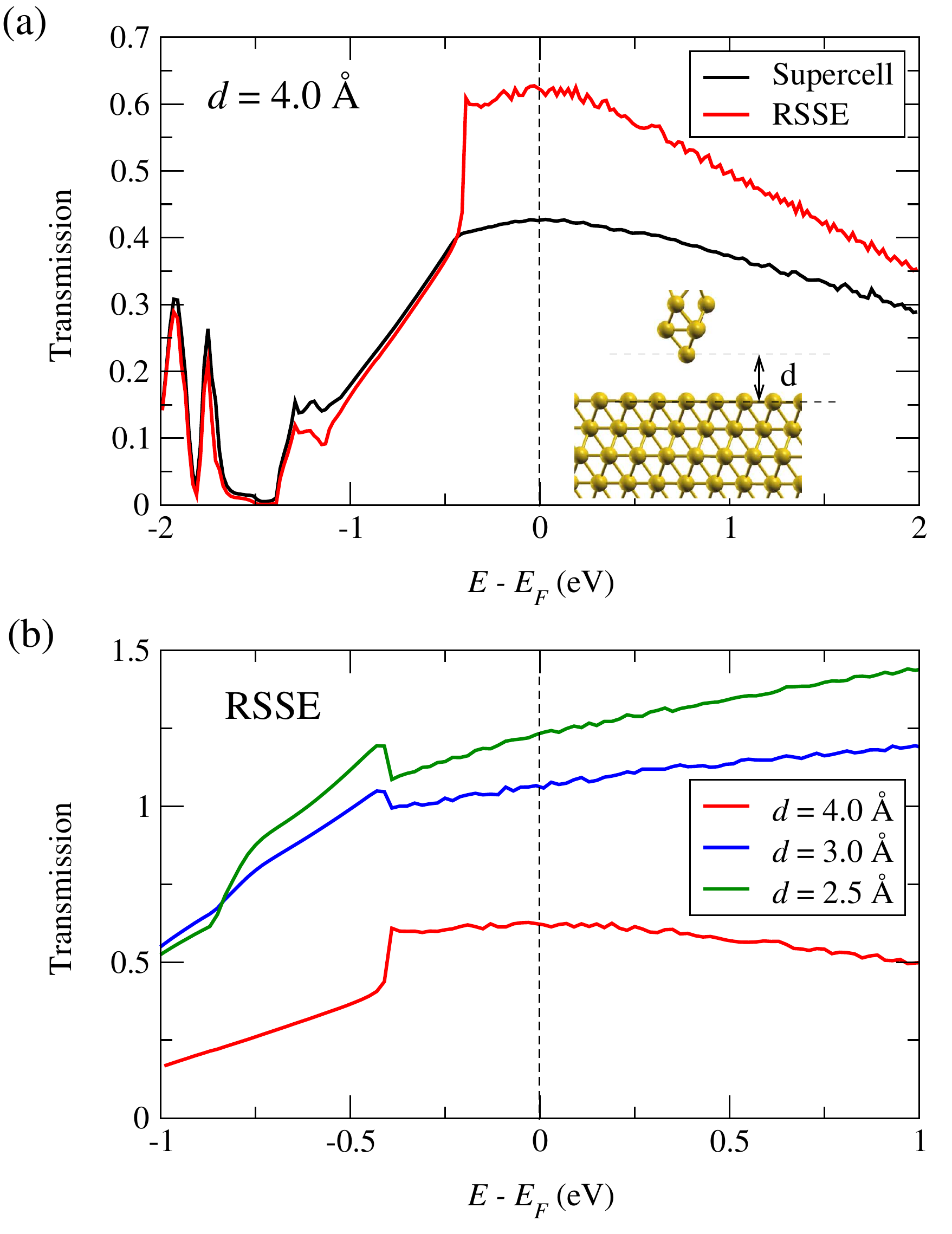}
	\caption{\label{au111} Clean Au(111): (a) Transmission functions from the tip to surface by supercell (black) and RSSE (red) approaches with $d = 4$ \AA. A strong increase of the transmission is seen with the RSSE approach at about $E_F-0.4\,\mathrm{eV}$ where the surface band appears. The supercell approach completely overlooks this contribution, resulting in over 30\% underestimation of the transmission coefficient near $E_F$. (b) RSSE transmissions at different $d$. The transmission shows a jump or a drop in the tunneling ($d = 4$ \AA) or contact ($d = 3$ \AA~or $d = 2.5$ \AA) regimes, respectively, at the SS band edge.
	}
\end{figure}

	We plot in Fig.~\ref{au111}a the transmission function from the tip into the clean Au(111) surface with a tip-surface distance ($d$) of 4 \AA. Most importantly, we find a crucial difference for energy higher than $-0.4\,\mathrm{eV}$ where the surface band appears and opens an additional transport channel. However, this surface channel is completely missing in the supercell approach. Therefore, this demonstrates the critical role of SS contributing more than 30\% to electron transmission around $E_F$. This is in very good agreement with the step-like feature observed in experiments \cite {Madhavan2001,Hollen_2015}. We further investigate the $d$ dependence on transmission, as shown in Fig. \ref{au111}b. Instead of an increase of the transmission at the SS band edge, we find a drop with decreasing $d$, and it more pronounced for smaller $d$. The transport through bulk and surface bands is independent in the tunneling regime while the two channels get coupled in the contact regime, which as we will argue later is at the origin of the transmission drop. 

	\begin{figure}[!t]
		\centering
		\includegraphics[width=1.0\linewidth]{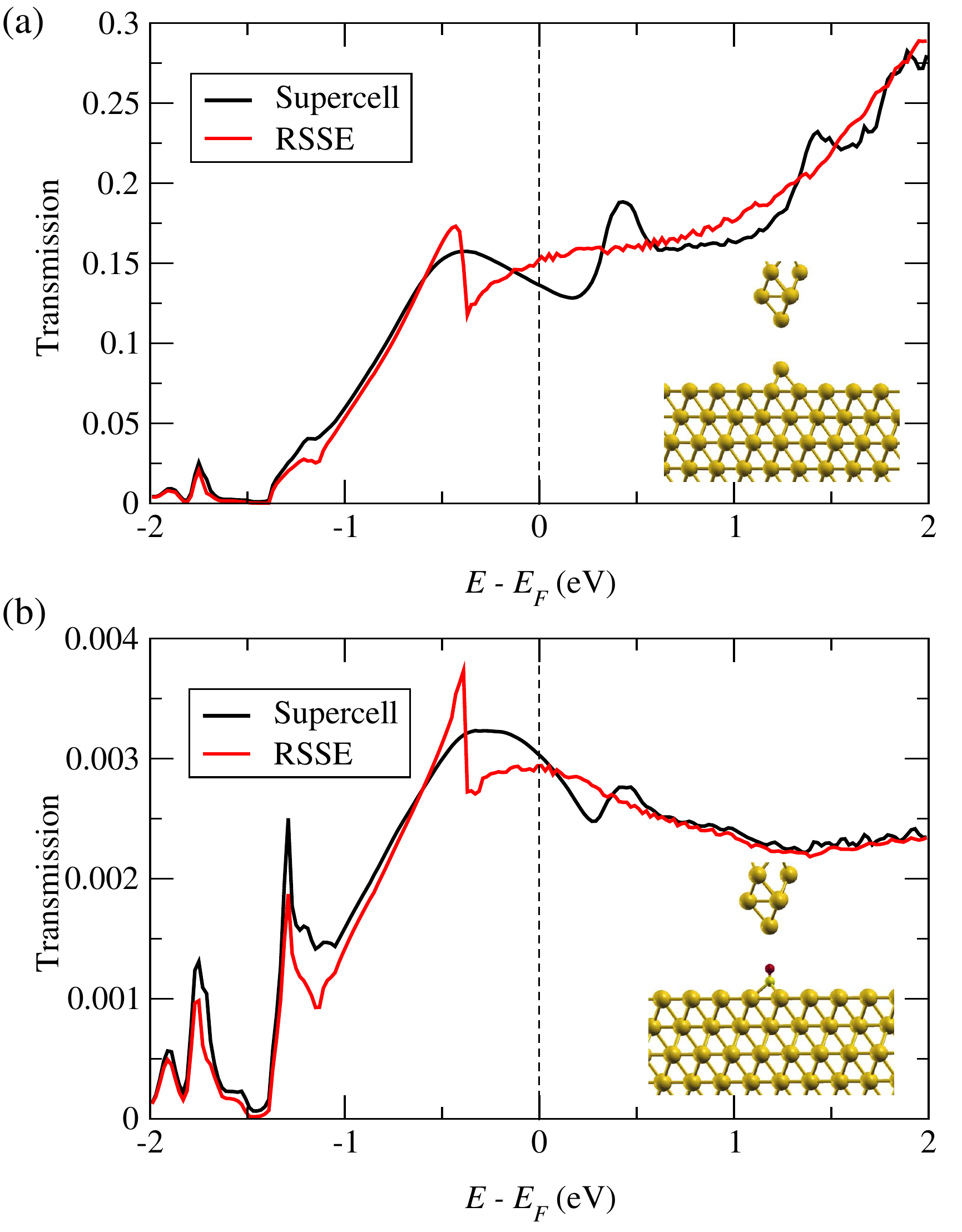}
		\caption{\label{adatom} Adatom and molecule on Au(111): The same as in Fig. \ref{au111}(a) but for adatom (a) and carbon monoxide molecule (b) deposited on surface. In the case of RSSE, a pronounced drop is found for both junctions at about $E_F-0.4\,\mathrm{eV}$. The distances between the tip and adatom or molecule are 4.5 \AA~and 3.6~\AA, respectively.}
	\end{figure}
	
	As another system, we consider a junction made of a single Au atom adsorbed in a hollow site on Au(111) with $d = 4.5\,\textup{\AA}$, as shown in Fig. \ref{adatom}a. Very interestingly, at about $E_F-0.4\,\mathrm{eV}$, we see again a transmission drop similar to the case of a clean surface. Due to the stronger interaction between adatom and surface, the drop becomes more pronounced. Our \emph{ab initio} results reproduce well the experimental measurements \cite{Olsson2004, Limot2005} in terms of energy level alignment and width of the peak. A similar transmission dip has also been recently reported in the case of Co adsorbed on Cu(111) by \textit{ab inito} theory \cite{tacca2021surface}. Moreover, the energy derivative of the transmission changes a sign in two methods leading to the opposite signs of the Seebeck coefficient which could be tested experimentally. We further note that this transmission drop is robust with respect to $d$ (see Appendix C). On the other hand, the supercell approach fails to reproduce the highly energy-dependent SS feature even with a very large lateral periodicity (see Fig. \ref{Fig_8}a in Appendix D). Finally, the Fano-factor (e.g., shot noise experiment) is very different when SS are included (see Fig. \ref{Fig_8}b in Appendix D). We argue that the transmission drop is a general effect resulting from the mixing of surface and bulk channels controlled by the interaction with, for example, an STM tip or an adsorbate. To verify this point, we investigated a carbon monoxide molecule adsorbed on a hollow site (energetically most favorable \cite{piccolo2004}) of the Au(111) surface as presented in Fig. \ref{adatom}b. Although the main conducting LUMO is located at about $E_F+1.8$ eV, very different from the gold adatom discussed before, a similar drop-like feature is again reproduced at the same energy, pointing out the general SS-related character.

	\begin{figure}[!t]
	\centering
	\includegraphics[width=1.0\linewidth]{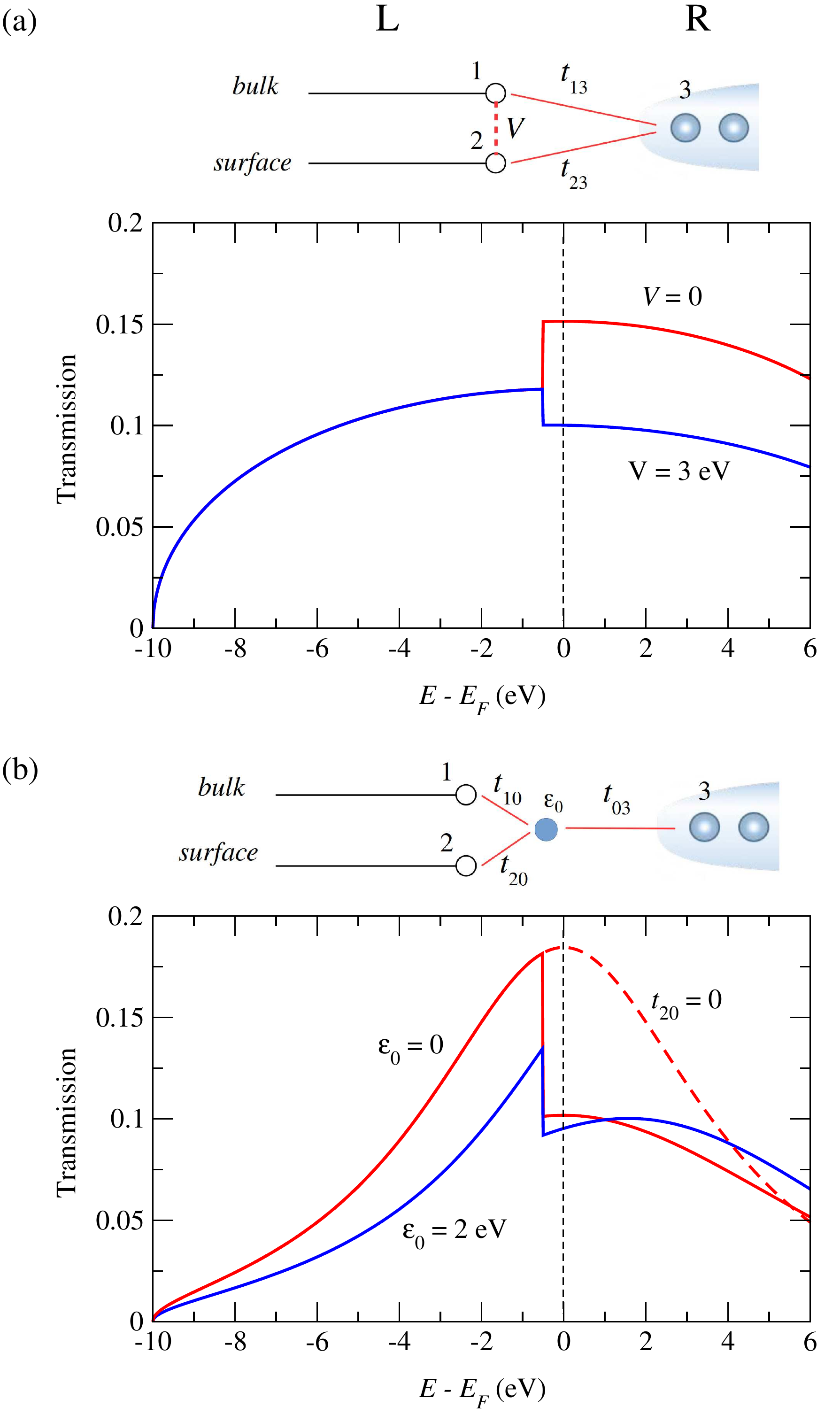}
	\caption{\label{model} Tight-binding model with two bands (left side), simulating the bulk and the surface channels of the Au(111) surface, and one band to the right, simulating the Au tip. (a) Model for clean Au(111) surface. The inclusion of interband coupling, $V$, allows to reproduce the transmission drop at the SS band edge; (b) Model for impurity on Au(111). The transmission drop is observed for any energy of impurity level but disappears if impurity/SS coupling is switched off (dashed red), confirming its QI interpretation.}
\end{figure}
	
	To understand the origin of the transmission drop at $-0.4\,\mathrm{eV}$ observed with the RSSE method, we construct a tight-binding (TB) model, depicted in Fig.~\ref{model}, with two bands for the electrode (left), simulating bulk (1) and surface (2) channels, and one band (3) for the STM tip (right) simulated by an atomic tight-binding chain with on-site and hopping energies being 0 and 5 eV, respectively.
	The surface (left) is characterized by unperturbed Green's functions at two contact orbitals, $g_1$ and $g_2$,  constructed on bulk and surface channels, respectively. They are chosen for simplicity to be purely imaginary (a wide-band approximation) as $g_i(E)=2\pi i/W_i$ for $|E-\varepsilon_i|<W_i$ and 0 otherwise. The band centers, $\varepsilon_i$, and half band-widths, $W_i$, are set to \{$\varepsilon_1=0, W_1=10~{\rm eV}$\} and 
	\{$\varepsilon_2=3.5~{\rm eV}, W_2=4~{\rm eV}$\} for bulk and surface bands, respectively, which gives in particular a SS band edge at $E=-0.5$ eV as required.
	
	We first consider the clean Au(111) contacted by the STM tip (Fig.~\ref{model}a). 
	The tip apex atom (3) is considered as a central region of Eq.~\ref{eq0}, the left side coupling $\Gamma_L$ is constructed on the $2 \times 2$ Green's function $G$ (from sites 1 and 2) of the surface while the right side $\Gamma_R$ is given by the tip TB chain expression. Necessary tip/surface coupling constants were set to $t_{13}=1$ eV and $t_{23}=0.35$ eV. If two channels of the surface are independent ($V=0$, so $G_{ij}=\delta_{ij}g_i$) the regular-shaped transmission is obtained with superposed contributions from bulk and surface bands (red). Interestingly, when the interband mixing $V$ is switched on, which corresponds to the increased bulk and SS mixing due to approaching STM tip, the transmission drop at the SS band edge is developed (blue). This drop can be attributed to destructive QI between the two electron pathways: i) the direct tunneling to the principal bulk channel, tip $\rightarrow$ bulk band (1) and ii) the pathway mediated by the SS, the tip $\rightarrow$ SS (2) $\rightarrow$ bulk band (1). Since both pathways involve the same terminal bulk band they can interfere destructively, producing the transmission drop. Overall the TB reproduces well the DFT results.
	
	For the impurity case, an additional orbital, $\epsilon_0$, should be introduced (Fig.~\ref{model}b). The $\Gamma_L$ for the tip apex orbital (3) is now constructed on the $3 \times 3$ Green's function $G$ (from sites 1, 2, and 0) whereas $\Gamma_R$ remains the same. The hopping parameters $t_{01}, t_{02}$, and $t_{03}$ were set to 5.0 eV, 3.0 eV, and 1.0 eV, respectively. It is assumed that the tunneling from the tip goes through the impurity orbital only without explicit inclusion of the interband mixing such that the two bands are coupled via the impurity level. The transmission drop appears again, independent of $\epsilon_0$. It is well reproduced, in particular, for $\epsilon_0=0$ (red) and $\epsilon_0=2$ eV (blue), which should correspond to Au adatom and the molecule, respectively, discussed before in Fig.~\ref{adatom}. Similar to the previous case, we can interpret this transmission drop to destructive QI due to two pathways: i) tip $\rightarrow$ impurity $\rightarrow$ bulk band and ii) tip $\rightarrow$ impurity $\rightarrow$ SS $\rightarrow$ impurity $\rightarrow$ bulk band. Note that if the impurity/SS coupling is turned off, $t_{20}=0$, the transmission recovers its normal shape (red dashed), which confirms our QI interpretation. 
	
	\begin{figure}[!t]
		\centering
		\includegraphics[width=1.0\linewidth]{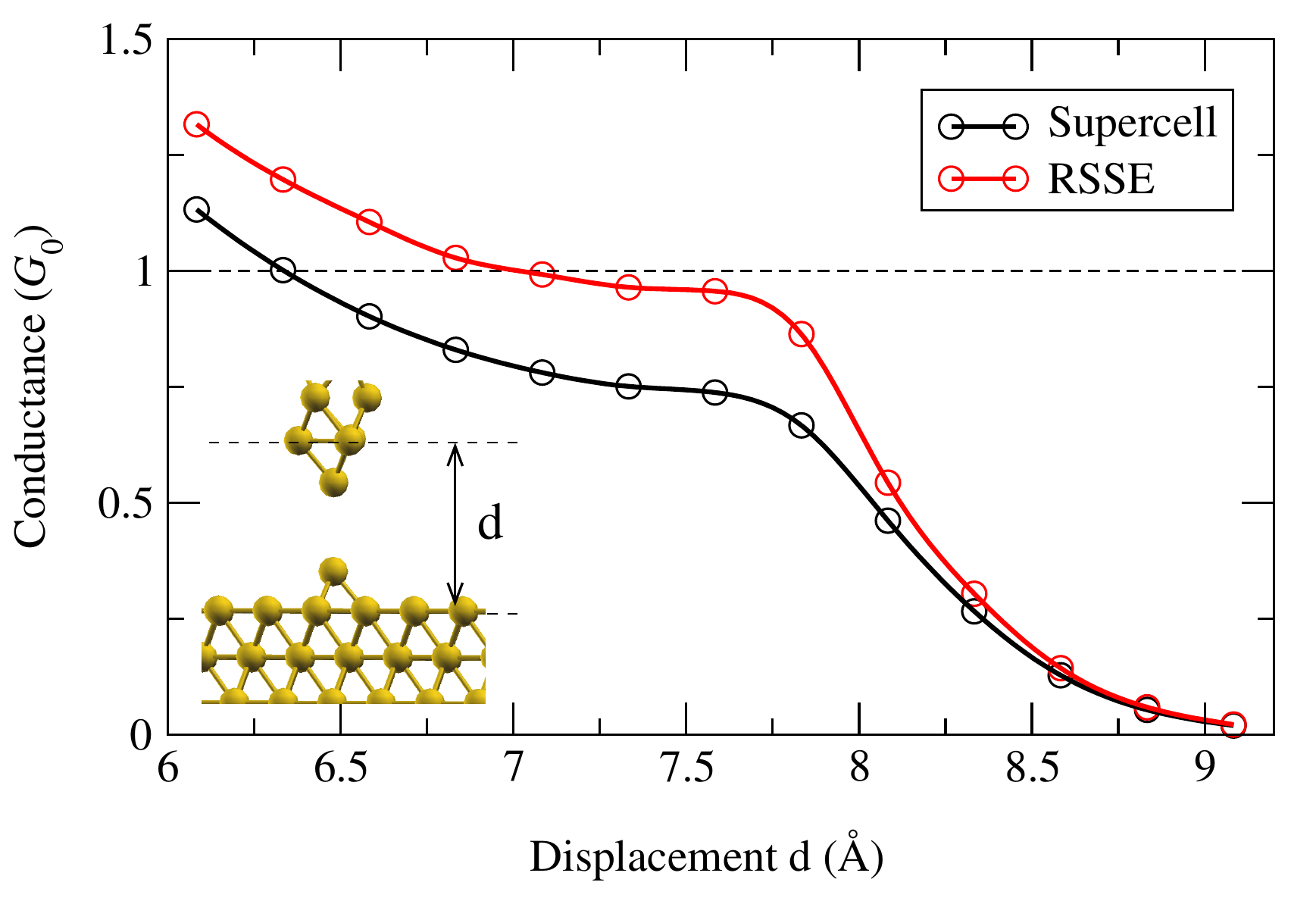}
		\caption{\label{cond} Conductance .vs. tip displacement for Au atomic contacts calculated by supercell (black) and RSSE (red) approaches. A conductance plateau of about 1~$G_0$ is found by the RSSE method, whereas the supercell approach yields only 0.8~$G_0$.}
	\end{figure}
	
	As mentioned before, the drop in transmission at the SS band edge is a robust feature showing up for different tip-adatom distances. As the tip is approaching the adatom, the SS are expected to give a more important contribution to the transmission function. We show, therefore, in Fig. \ref{cond} the conductance as a function of $d$. Clearly, close to the contact regime ($d<8$~\AA), the difference between the RSSE and conventional supercell approach starts to increase, reflecting the enhanced contribution of SS into the transport. Due to SS contribution, missing in the supercell approach, the RSSE method provides much better quantized electron transport with the conductance of $1\,G_0$ improving strongly an agreement with experimental data \cite{Brandbyge1995,Vardimon2013,chen2014}. Being generic, the conductance jump at the edge of the SS should be robust in STM setup but less pronounced in break junction setup where the SS are strongly perturbed by a pyramid-like cluster formed on the surface.

\section{CONCLUSIONS}
	
	In summary, using the newly developed RSSE method, we have demonstrated how surface states can be properly incorporated in \textit{ab initio} electron transport going beyond current state-of-the-art DFT+NEGF. We show that the SS can contribute more than 30\% in electron transport near $E_F$ for noble metal surfaces. Moreover, we find a robust QI induced by SS in both atomic and molecular junctions ranging from tunneling to contact regime, in good agreement with experimental measurements. We also find that the RSSE gives a considerably better quantized conductance for metallic nanocontacts. Our method should be important for a variety of systems where surface propagating channels can play an important role, such as topologically derived SS \cite{seo2010,yan2015}, graphene edge states \cite{Yao2009}, and 2D materials/metal interfaces \cite{dahal2014}.
	
	\noindent{\bf Acknowledgment}
	We acknowledge funding from the EU Horizon 2020 programme (766726), Villum foundation (13340), and the Danish Research Foundation (DNRF103). Part of the calculations was done using HPC resources from CALMIP (P21023).
	
	
	\bibliographystyle{apsrev}
	\bibliography{References}

	\section*{Appendix A: Computational details}
	\label{Appendix-A}

\begin{figure}
	\includegraphics[scale=0.55]{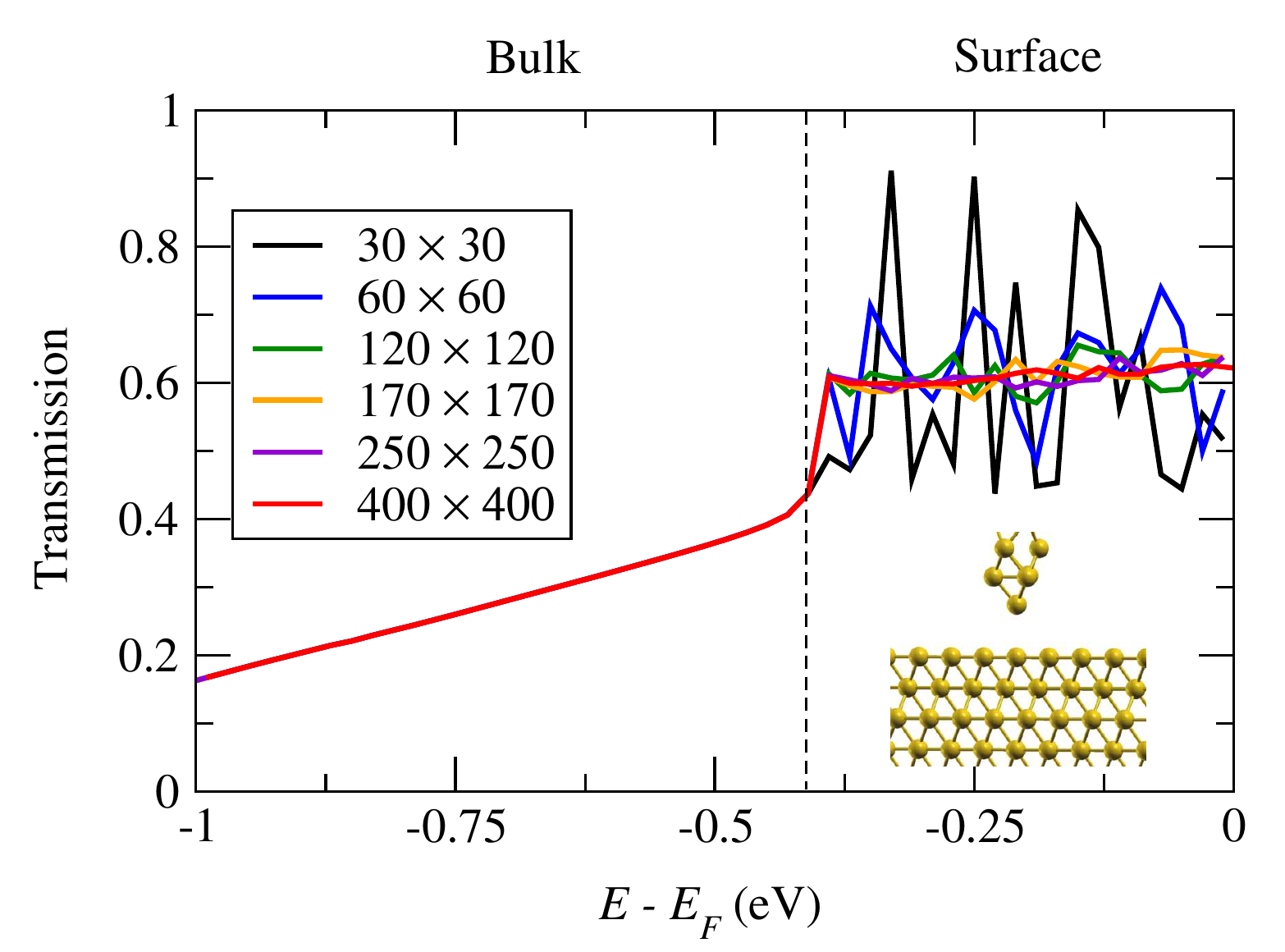}
	\caption{\label{Fig_6} The convergence of the self-energy with respect to planar 2D $\V{k}$ points in $6 \times 6$ unit cell. The surface states' transmission, at $E>-0.4$ eV, converges much slower compared to the bulk one.}
\end{figure} 

$\textit{Ab initio}$ quantum transport calculations were done using $\textsc{TranSIESTA}$ \cite{Brandbyge-2002,papior2017improvements} in the LDA with the parametrization of Perdew and Zunger \cite{Perdew1981} for exchange-correlation functional. The valence–electron wavefunctions were expanded in a basis set of local orbitals in $\textsc{SIESTA}$ \cite{soler2002siesta,garcia2020siesta}. Concerning the basis set, the cutoff radius ($r_c$ in Bohr units) and orbitals are $r_c$ = 6.083, 5.715 (6$s$, polarized) and $r_c$ = 4.287, 2.838 (5$d$) for the Au. The pseudopotentials were taken from Ref.~\onlinecite{rivero2015} where the parameters have been carefully checked, producing a good agreement with plane-wave $\textit{ab initio}$ code. The mesh cutoff and filter cutoff were set to 400 Ry and 300 Ry, respectively. The lattice constants of the Au were set according to their bulk experimental lattice constants of 4.07~\AA~and the interlayer spacing within the slab was held fixed. The spin-orbit coupling was not considered. For the conductance evaluation with respect to displacement distance (see Fig. \ref{cond}), we allow only STM apex and adatom atoms to relax until atomic forces became lower than 0.01 eV/\AA, and repeat it until forming atomic contacts. Six atomic layers were used to describe the surface with bulk self-energy on the lower three. All the results presented were calculated within a large supercell of $6 \times 6$ Au(111) unit cells for both supercell and RSSE methods. 

In the supercell approach, the density matrix was converged using $6\times6$ \kpts~while the transmission functions were evaluated using denser $24\times24$ \kpts. For the RSSE method, the self-energy was constructed using a very dense $\V{k}$-mesh of $400\times400$ and the transmission spectrum was calculated within a single \kpts. Physical quantities like transmission, the PDOS, and Fano-factor were extracted using TBtrans and SISL \cite{zerothi_sisl}.

\section*{Appendix B: $\V{k}$ points convergence}
	\label{Appendix-B}
	
For the pre-calculation of RSSE, we need a dense integration grid in reciprocal space to obtain well-converged surface states' transmission as shown in Fig. \ref{Fig_6}, at $E>-0.4$ eV. The bulk contribution, $E<-0.4$ eV, converges very fast (even with $30 \times 30$) while much denser $\V{k}$ mesh is needed for the surface states. Throughout the paper, the real-space self-energy was calculated from $400 \times 400$ $\V{k}$ points.
	
	\section*{Appendix C: From tunneling to contact regimes}
		\label{Appendix-C}
	
\begin{figure}
	\includegraphics[scale=0.52]{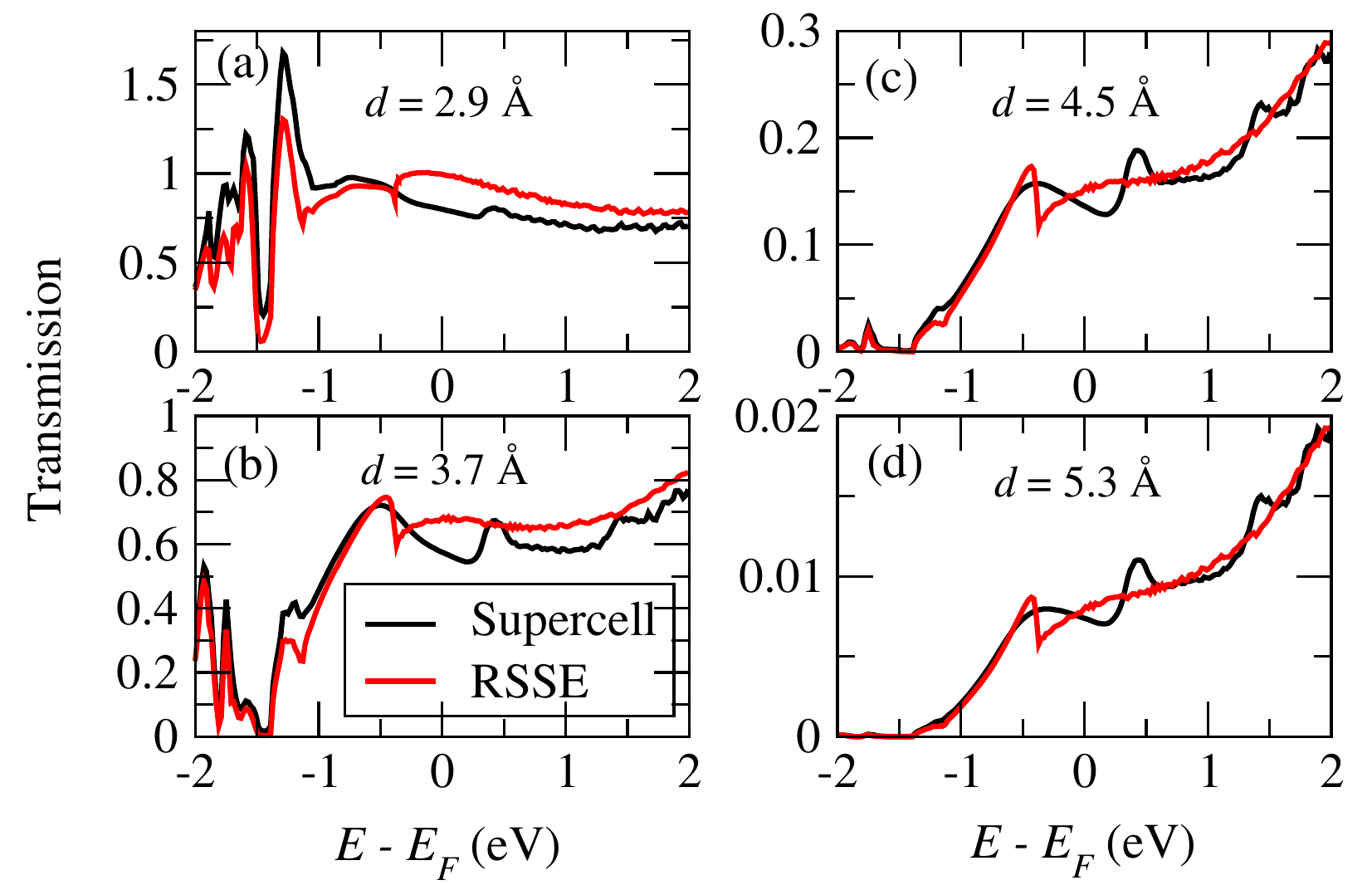}
	\caption{\label{Fig_7} Adatom on Au(111): Transmission functions calculated by two different approaches with different tip-adatom distances ranging from tunneling to contact regime: (a) $d=2.9$ \AA, (b) $d=3.7$ \AA, (c) $d=4.5$ \AA, and (d) $d=5.3$ \AA.}
\end{figure}

The destructive quantum interference is found to be rather robust, appearing for different tip-atom distances ranging from tunneling to contact regime as Fig. \ref{Fig_7} shows. Note that the smaller the tip-adatom distance, the larger contribution of surface states around the Fermi energy. In particular, at the contact regime ($d=2.9$ \AA), the transmission calculated with our RSSE approach starts to increase at the surface band edge ($E>-0.4$ eV) 
deviating considerably from the one obtained with the conventional supercell method. As a consequence, with RSSE we recover the conductance very close to the quantized value of 1~$G_0$, often reported in single-atom contact measurements. 

	\section*{Appendix D: Periodicity size effect for the supercell approach}
	\label{Appendix-D}

\begin{figure}[!htbp]
	\includegraphics[width=0.9\linewidth]{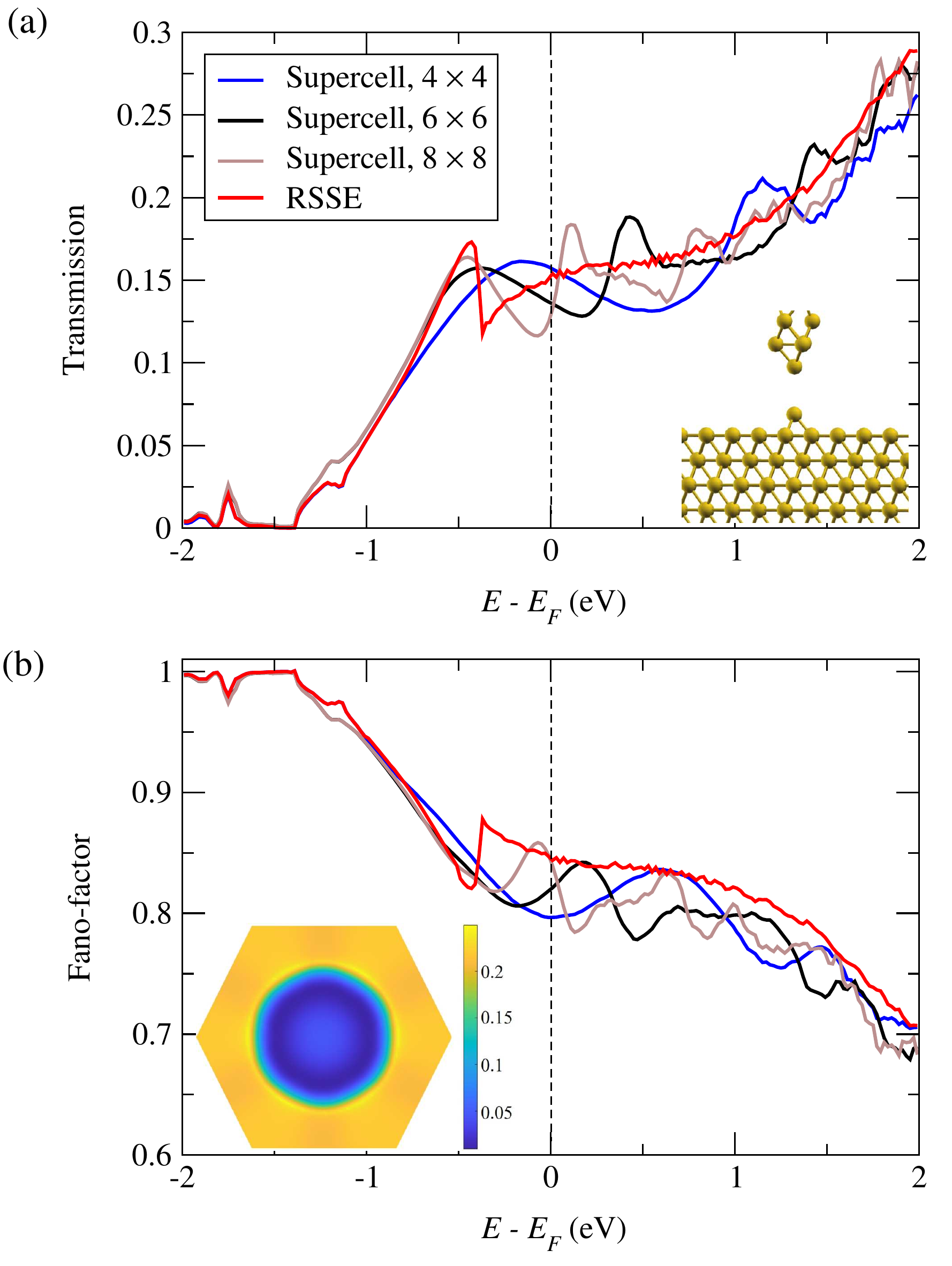}
	\caption{\label{Fig_8} Adatom on Au(111): Transmission functions (a) and Fano-factor (b) calculated by RSSE (red) and the supercell approach with different in-plane periodicity, $4 \times 4$ (blue), $6 \times 6$ (black) and $8 \times 8$ (brown). For supercell approaches, standing-wave patterns are found and they become denser when the periodicity is increased. The \Vk-dependence on transport is also reflected in $\V{k}_{\parallel}$-resolved transmission coefficients at the Fermi level. The tip atom to adatom distance is 4.5 \AA.}
\end{figure}

	To investigate how the in-plane periodicity influences the transport behavior, we plot in Fig.~\ref{Fig_8}a the transmission functions calculated by the supercell approach with different sizes of the periodic array. Even for the relatively wide $8 \times 8$ superlattice (434 atoms in total), a large difference is found compared to the RSSE result. Interestingly, for the PBCs, we find standing wave-like patterns in the SS energy window which become denser when the in-plane periodicity is increased, similar to confined states in a quantum well of increasing width. Even if the size of the supercell could be made sufficiently large, while likely computationally unfeasible, we would {\em not} necessarily end up with the RSSE result, drawn with the red line, which lacks these oscillations. This is due to the fact that for PBCs it is only the asymptotic bulk scattering states which are represented in the coupling matrices, $\V{\Gamma_{\text{L/R}}}$, entering the transmission. On the other hand, the adsorbate atom potential may provide a coupling between surface and bulk states, decreasing the importance of the asymptotic SS channel. 

 On the other hand, fig.~\ref{Fig_8}b shows the corresponding Fano-factor calculated from $\V{k}$-dependent $i$-th transmission eigenchannels, $\tau_{i,\V{k}}$. 
\begin{equation}
\label{eq3}
F=\frac{\sum_{i,\V{k}}\tau_{i,\V{k}}(1-\tau_{i,\V{k}})\omega_\V{k}}{\sum_{i,\V{k}}\tau_{i,\V{k}}\omega_\V{k}}
\end{equation}
where $\omega_\V{k}$ is \kpts\ weight-factors. Due to the PBCs and \kpt\ sampling, the Fano-factor is sensitive to the periodic repetition. This is also reflected in $\V{k}_{\parallel}$-resolved transmission coefficients at the Fermi level in first-Brillouin. The use of RSSE removes the artificial interactions between the periodic images and $\V{k}$-dependence. 

\end{document}